\documentclass[10pt, twocolumn, amssymb, aps, prx, reprint, floatfix, superscriptaddress]{revtex4-2}
\usepackage[a4paper, total={7in, 10in}]{geometry}
\usepackage{graphicx} 
\graphicspath{{Figures/}}
\usepackage{amsmath}
\usepackage{physics}
\usepackage{dsfont}
\usepackage{dcolumn}
\usepackage{xcolor}
\usepackage{bbold}
\usepackage{appendix}
\usepackage{units}

\usepackage{pgfplots}
\pgfplotsset{compat=1.18} 

\usepackage{hyperref}
\hypersetup{colorlinks=true,linkcolor=blue}

\begin{document}

\title{Multi-level spectral navigation with geometric diabatic-adiabatic control}

\author{Christian Ventura-Meinersen}
\thanks{Contact author: c.venturameinersen@tudelft.nl}
\affiliation{QuTech and Kavli Institute of Nanoscience, Delft University of Technology, PO Box 5046, 2600 GA Delft, The Netherlands
}

\author{Edmondo Valvo}
\affiliation{QuTech and Kavli Institute of Nanoscience, Delft University of Technology, PO Box 5046, 2600 GA Delft, The Netherlands
}

\author{Stefano Bosco}
\affiliation{QuTech and Kavli Institute of Nanoscience, Delft University of Technology, PO Box 5046, 2600 GA Delft, The Netherlands
}

\author{Maximilian Rimbach-Russ}
\affiliation{QuTech and Kavli Institute of Nanoscience, Delft University of Technology, PO Box 5046, 2600 GA Delft, The Netherlands
}

\date{\today}

\begin{abstract}
    We introduce a geometric framework for efficient few-parameter pulse optimization in multi-level quantum systems, enabling high-fidelity state transfer beyond the adiabatic limit. Our method interpolates smoothly between adiabatic and diabatic dynamics to minimize unwanted excitations and maximize desired transitions even within a multi-level structure. Crucially, for single-parameter pulse control, the optimization reduces to solving a first-order ordinary differential equation. We showcase the flexibility of our diabatic-adiabatic protocols through two examples in spin-based quantum information processing: state initialization and qubit state transfer.
\end{abstract}

\maketitle

\textit{Introduction --- }A universal challenge for the control of any multi-level quantum system is the efficient navigation of quantum states through complex parameter space landscapes~\cite{glaserTrainingSchrodingersCat2015}. Dense spectra occur in various material platforms from superconducting circuits~\cite{izmalkovObservationMacroscopicLandauZener2004, berkeTransmonPlatformQuantum2022}, trapped-ions~\cite{mosesRaceTrackTrappedIonQuantum2023}, photonic circuits~\cite{alexanderManufacturablePlatformPhotonic2025}, and many more~\cite{higuchiLightfielddrivenCurrentsGraphene2017, aghaeeInterferometricSingleshotParity2025, pettaCoherentBeamSplitter2010,ribeiroCoherentAdiabaticSpin2013, binderQuantacellPowerfulCharging2015}. A wide range of approaches have been developed to improve performance and control. These include mathematical control theory~\cite{khanejaTimeOptimalControl2001, carliniTimeOptimalQuantumEvolution2006,pontryaginMathematicalTheoryOptimal2017,boltyanskiGeometricMethodsOptimization1999,carliniTimeoptimalUnitaryOperations2007,boscainIntroductionPontryaginMaximum2021,koikeQuantumBrachistochrone2022, rezakhaniQuantumAdiabaticBrachistochrone2009,walelignDynamicallyCorrectedGates2024, stepanenkoTimeoptimalTransferQuantum2025}, implementing advanced numerical methods~\cite{khanejaOptimalControlCoupled2005,goodwinAcceleratedNewtonRaphsonGRAPE2023, canevaChoppedRandombasisQuantum2011, fauquenotOpenClosedLoop2025, katiraee-farUnifiedEvolutionaryOptimization2025}, spectral approaches~\cite{martinisFastAdiabaticQubit2014, rimbach-russSimpleFrameworkSystematic2023, wuSimultaneousHighFidelitySingleQubit2025, polatPulseShapingUltraFast2025}, and employing techniques designed to accelerate adiabatic processes, dubbed shortcut-to-adiabaticity~\cite{bergmannCoherentPopulationTransfer1998,motzoiSimplePulsesElimination2009,berryTransitionlessQuantumDriving2009,degrandiAdiabaticPerturbationTheory2010,ivakhnenkoNonadiabaticLandauZener2023,vuthaSimpleApproachLandauZener2010,motzoiOptimalControlMethods2011,chenLewisRiesenfeldInvariantsTransitionless2011,ribeiroSystematicMagnusBasedApproach2017,selsMinimizingIrreversibleLosses2017,theisCounteractingSystemsDiabaticities2018,banFastLongrangeCharge2018,takahashiHamiltonianEngineeringAdiabatic2019,banSpinEntangledState2019,guery-odelinShortcutsAdiabaticityConcepts2019,setiawanAnalyticDesignAccelerated2021,takahashiDynamicalInvariantFormalism2022,glasbrennerLandauZenerFormula2023,dengisAcceleratedCreationNOON2025,romeroOptimizingEdgestateTransfer2024, liuAcceleratedAdiabaticPassage2024,xuImprovingCoherentPopulation2019,fehseGeneralizedFastQuasiadiabatic2023,limaPartialLandauZenerTransitions2025,richermeExperimentalPerformanceQuantum2013, rolandQuantumSearchLocal2002,martinez-garaotFastQuasiadiabaticDynamics2015,chenSpeedingQuantumAdiabatic2022, fernandez-fernandezFlyingSpinQubits2024, baksicSpeedingAdiabaticQuantum2016}. 

An approach to gain operational performance with enhanced interpretability exploits the geometry of quantum dynamics, as those based on geometric space curves~\cite{barnesDynamicallyCorrectedGates2022, zhuangNoiseresistantLandauZenerSweeps2022, wangCompositePulsesRobust2012, piliourasAutomatedGeometricSpace2026} and the quantum metric tensor~\cite{ventura-meinersenQuantumGeometricProtocols2025, meinersenUnifyingAdiabaticStatetransfer2025, liskaHiddenSymmetriesBianchi2021}. 
A key requirement to efficiently manipulating quantum states is to perform targeted excitations, beyond the adiabatic limit with minimal overhead. Employing piecewise-constant sudden-switch methods, which locally approximate the energy spectrum by a two-level Landau-Zener model, one can engineer transitions at the cost of a discontinuous pulse~\cite{murgidaCoherentControlInteracting2007, murgidaCoherentControlLocalization2009}. With the rise of larger quantum systems, there is a need to find a framework that smoothly interpolates between adiabatic and diabatic dynamics during the time evolution, to allow for experimental feasibility with finite bandwidth, without requiring approximations or constraints to the energy spectrum.

\begin{figure}[t!]
    \centering
    \includegraphics[width=\linewidth]{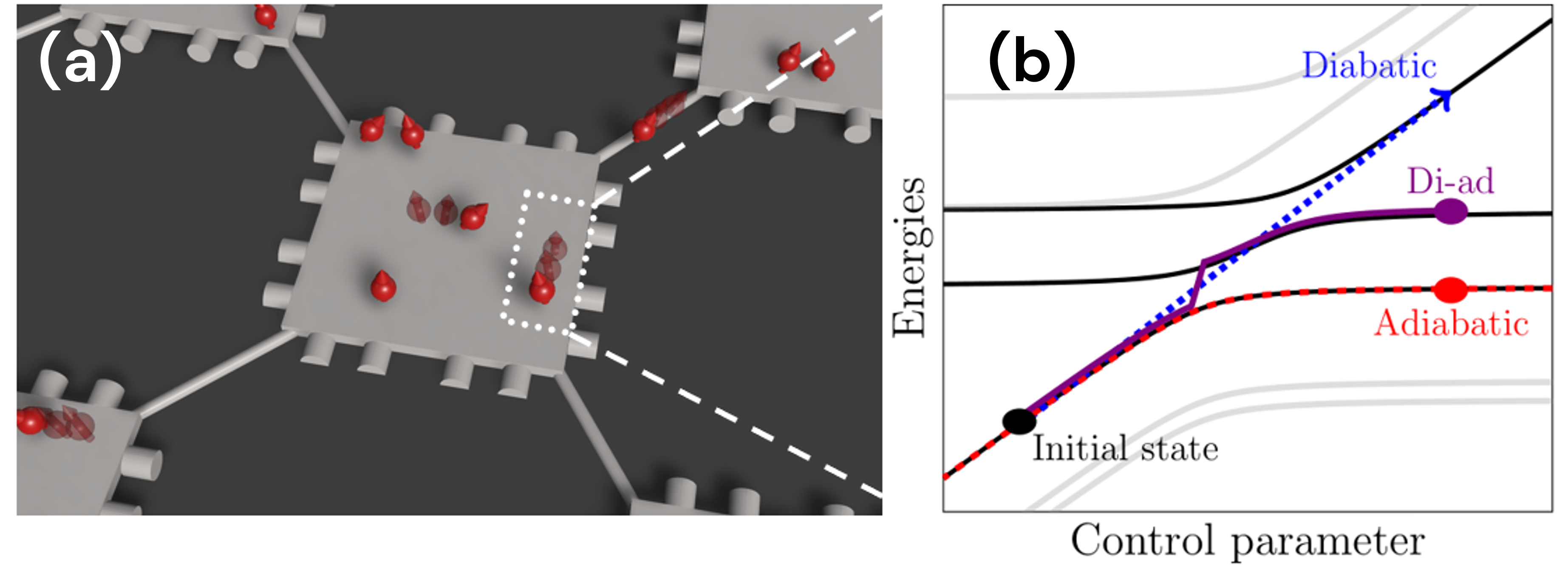}
    \caption{Application of the di-ad protocol \textbf{(a)}, here, spin qubit shuttling, and the emergent dense energy spectrum as a function of some control parameter \textbf{(b)}. The di-ad protocols allow for flexible and high-fidelity state transfer beyond the extrema of adiabatic (red, dashed) and diabatic (blue, dashed) limits by selectively operating between both regimes (purple).}
    \label{fig: diad concept}
\end{figure}

In this Letter, we present a versatile approach to perform efficient single-parameter pulse shaping of experimentally accessible parameters of the system's Hamiltonian  making it ideal for any multi-level quantum system. The single-parameter pulse shaping efficiency is granted because the resulting optimization task maps to a simple first-order ordinary differential equation. Our resulting framework provides various diabatic-adiabatic (di-ad) protocols (See Fig.~\ref{fig: diad concept}), which are described by a continuous family of Riemannian geometries. The di-ad protocols allow for enhanced experimental flexibility, as one can adjust the pulse smoothness to the experimental hardware constraints by continuously deforming the resulting Riemannian geometry to one that is experimentally achievable. Henceforth, our geometric framework presents a high-fidelity, experimentally-accessible, system-agnostic approach to optimal quantum state control.    

We illustrate the applicability of the di-ad protocol by studying two exemplary open problems in spin-based quantum information processing: quantum state initialization and quantum state shuttling.

\textit{Optimal control with di-ad tensor --- }The task of quantum information transfer can be stated as a dynamical problem, where a given initial state $\ket{\psi(0)}$ is mapped to a target state $\ket{\psi_\text{target}}=\ket{\psi(t_\text{f})}$,  with $t_\text{f}$ being the pulse time,  via engineered time evolution. Here we resort to a geometric formulation using the quantum metric tensor~\cite{ventura-meinersenQuantumGeometricProtocols2025,meinersenUnifyingAdiabaticStatetransfer2025, liskaHiddenSymmetriesBianchi2021}, where the above task is framed as the geodesic evolution on the space of experimentally-accessible parameters. Given a set of control parameters $x^\mu$ of the system Hamiltonian $\hat{H}[x^\mu]$ we want to map an initial state $\ket{\psi(0)}=\ket{\psi[x^\mu(0)]}$ to the final target state $\ket{\psi_\text{target}}=\ket{\psi[x^\mu(t_\text{f})]}$ through the geodesic evolution of $x^\mu(\tau)$, where $\tau=t/t_\text{f}$ is the rescaled time parameter. The infinitesimal state overlap of the evolved state can be expressed by the quantum metric tensor $g_{\mu\nu}(x)$ ~\cite{liskaHiddenSymmetriesBianchi2021} 
\begin{align}
    1-|\braket{\psi(x)}{\psi(x+dx)}|^2\approx g_{\mu\nu}(x)dx^\mu dx^\nu.
\end{align}
In the eigenbasis of the Hamiltonian and in the adiabatic limit, the quantum metric tensor is related to the energy fluctuations~\cite{chengQuantumGeometricTensor2013}
\begin{align}
    \label{eqn: energy fluctuations qmt}
    \sigma_E^2=\expval{\hat{H}^2}-\expval{\hat{H}}^2=g_{\mu\nu}(x)\dv{x^\mu}{\tau}\dv{x^\nu}{\tau},
\end{align}
where $g_{\mu\nu}$, with respect to a given state $\ket{\psi_m}$, can then be conveniently written in terms of the Hamiltonian $\hat{H}$, its eigenvalues and eigenvectors $\{E_n,\ket{\psi_n}\}$
\begin{align}
    \label{eqn: qgt Hamiltonian}
    g_{\mu \nu}=\Re \sum_{n\neq m} \frac{\mel{\psi_m}{\partial_\mu \hat{H}}{\psi_n}\mel{\psi_n}{\partial_\nu \hat{H}}{\psi_m}}{(E_n-E_m)^2},
\end{align}
where $\partial_\mu = \partial/\partial x^\mu$ is the derivative with respect to the parameters of the Hamiltonian. If we assume that the energy fluctuations are constant, we can minimize the infinitesimal state overlap by minimizing Eq.~\eqref{eqn: energy fluctuations qmt}~\cite{ventura-meinersenQuantumGeometricProtocols2025}. Finding the optimal control pulse that minimizes the energy fluctuations is equivalent to solving the geodesic equations. The quantum metric depends on two fundamental aspects of the control Hamiltonian: the overlap matrix elements of its gradients $\mel{\psi_m}{\partial_\mu \hat{H}}{\psi_n}$, with respect to the control parameter $x^\mu$,  and the corresponding energy splittings $(E_n-E_m)$ of the coupled eigenstates. Together, these factors determine the resulting pulse shape and the associated state transfer infidelity. For adiabatic state transfer, this protocol achieves high fidelities~\cite{ventura-meinersenQuantumGeometricProtocols2025}, yet it is not well-suited for diabatic evolution.  Deducing from the adiabatic impulse approximation~\cite{tomkaAccuracyAdiabaticimpulseApproximation2018, suzukiGeneralizedAdiabaticImpulse2022}, we now locally maximize the components of the quantum metric to achieve diabatic state transfer~\cite{sm}. Despite the non-perturbative nature of diabatic evolution, we make a phenomenological ansatz by defining a generalized quantum metric tensor
\begin{align}
    \mathcal{G}_\text{di-ad}&\equiv\left(\mathcal{G}_\text{di-ad}\right)_{\mu\nu}^{(\alpha,\beta;\;\hat{\alpha},\hat{\beta})} \\
    &=\mathcal{A}_{\mu\nu}^{(\alpha,\beta)}+\mathcal{D}_{\mu\nu}^{(\hat{\alpha},\hat{\beta})} \notag \\
    &=\sum_m\sum_{n\neq m} \Big([1-\xi_{mn}]\,\mathcal{G}_{\mu\nu}^{\,nm,(\alpha,\beta)}
    +\xi_{mn}\,\mathcal{G}_{\mu\nu}^{\,nm,(\hat{\alpha},\hat{\beta})}\Big), \notag
\end{align}
which we call the \textit{di-ad tensor}. The di-ad tensor is comprised of an adiabatic component $\mathcal{A}$ and a potentially diabatic component $\mathcal{D}$. The individual components read identically, but may include different exponents,
\begin{align}
    \mathcal{G}_{\mu\nu}^{\,nm,(\alpha,\beta)} = \Re \frac{\mel{\psi_m}{\partial_\mu \hat{H}}{\psi_n}^{\beta/2}\mel{\psi_n}{\partial_\nu \hat{H}}{\psi_m}^{\beta/2}}{|E_n-E_m|^\alpha}
\end{align}
and the components of the \textit{di-ad transition matrix} $\xi_{mn}\in\{0, 1\}$ are defined in the eigenbasis of the Hamiltonian as 
\begin{align}
    \xi_{mn}=\begin{cases}
        0 & \text{adiabatic: } \ket{\psi_m}{\to}\ket{\psi_m} \text{ and }\ket{\psi_m}{\not\to}\ket{\psi_n}\\
        1 & \text{diabatic: } \ket{\psi_m}{\to}\ket{\psi_n}
    \end{cases}
\end{align}
Via the di-ad transition  $\xi_{mn}$, one is only required to establish the initial and final target states, even if they are not adiabatically connected. The parameters $(\alpha,\beta;\; \hat{\alpha}, \hat{\beta})$ determine the degree of adiabaticity and diabaticity, respectively, as we will now see. For sufficiently separated level crossings, one can approximate a two-level crossing by a local Landau-Zener model~\footnote{Adding a $\sigma_y$ term allows us to rewrite the Hamiltonian as $\hat{H}=\vec{n}\cdot \vec{\sigma}$ with the unit vector on the sphere $\vec{n}$. The resulting geometry is then given by the geometry of the Bloch sphere~\cite{meinersenUnifyingAdiabaticStatetransfer2025}. For state transfer between the computational states $\ket{0},\ket{1}$, we only require the evolution along the azimuthal angle, which corresponds to $z(t)$ in the main text.} $\hat{H}_\text{LZ}/x=z(t)/x\,\sigma_z + \sigma_x$, with control parameter $z(t)$, one finds that the di-ad tensor simplifies to 
\begin{align}
    \begin{split}
        \mathcal{G}_\text{di-ad} = [1-\xi_{01}]\,\mathcal{G}_{zz}^{\,01,(\alpha,\beta)}+\xi_{01}\,\mathcal{G}_{zz}^{\,01,(\hat{\alpha},\hat{\beta})},
    \end{split}
\end{align}
where the di-ad transition matrix is now determined, whether we want to have adiabatic ($\ket{\psi_0}{\mapsto} \ket{\psi_0}$, $\xi_{01}{=}0$) or diabatic ($\ket{\psi_0}{\mapsto} \ket{\psi_1}$, $\xi_{01}{=}1$) evolution, respectively. The di-ad tensor for the Landau-Zener model can be computed analytically and reads $\mathcal{G}_\text{di-ad}{\simeq}(1+z(t)/x)^{-n_+}$, where $n_+{=}(\alpha{+}\beta)/2$ ($n_+{=}(\hat{\alpha}{+}\hat{\beta})/2$) labels all the different adiabatic (diabatic) pulses. In this simple case, the di-ad tensors for both evolutions are identical and only depend on $n_+$. In more complex systems, each parameter $(\alpha,\beta; \, \hat{\alpha}, \hat{\beta})$ will provide a unique contribution. Using the di-ad tensor, we can extend the geometric-adiabatic condition in Eq.~\eqref{eqn: energy fluctuations qmt} by substituting the standard quantum metric tensor with the di-ad tensor $g_{\mu\nu}\mapsto \mathcal{G}_\text{di-ad}$. The pulse shape is then given by solving for the appropriate differential equation~\cite{sm} labelled by $n_+$. 

\begin{figure}[t!]
    \centering
    \includegraphics[width=\linewidth]{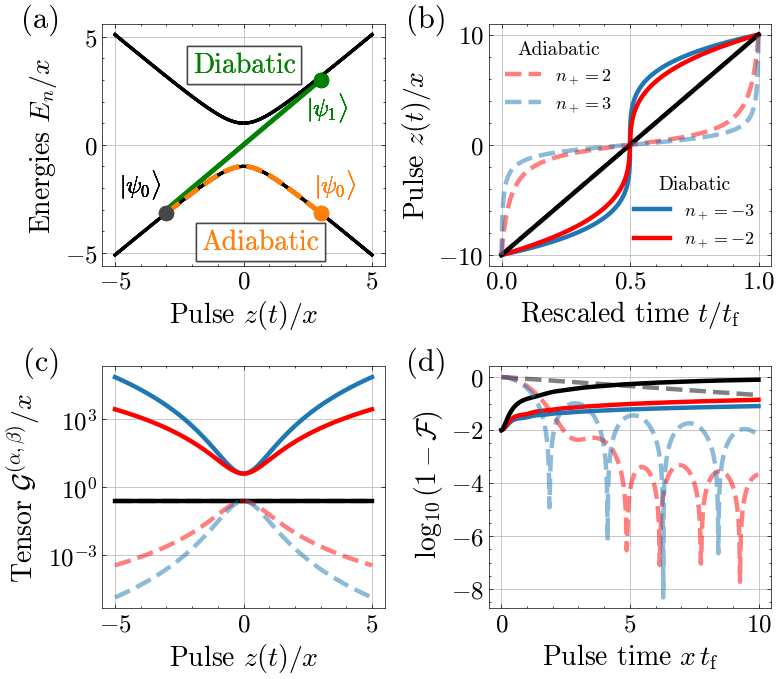}
    \caption{Coherent state transfer for the adiabatic ($\ket{\psi_0}\mapsto \ket{\psi_0}$, dashed lines) and the diabatic ($\ket{\psi_0}\mapsto \ket{\psi_1}$, solid lines) evolution in the Landau-Zener model. We show the energy spectrum \textbf{(a)}, the di-ad pulse \textbf{(b)}, originating from the corresponding di-ad metric \textbf{(c)}, and the resulting infidelities $1{-}\mathcal{F}$ \textbf{(d)} with $\mathcal{F}{=}|\braket{\psi_0(t_\text{f})}{U(t_\text{f})\psi_0(0)}|^2$ for adiabatic and $\mathcal{F}{=}|\braket{\psi_1(t_\text{f})}{U(t_\text{f})\psi_0(0)}|^2$ for diabatic evolution, where $U(t_\text{f})$ describes the unitary evolution given the pulse $z(t)/x$. The Hamiltonian parameters used are $z(0)/x{=}{-}z(t_\text{f})/x{=}{-}10$. The colored pulses refer to different $n_+$ with the appropriate sign for adiabatic (dashed) and diabatic (solid) protocols: $|n_+|=0,2,3$  corresponds to black, red, and blue, respectively.}
    \label{fig: LZ}
\end{figure}

In Fig.~\ref{fig: LZ}, we illustrate the possible state transfer protocols for the Landau-Zener problem and find that pulses with $n_+>0 \,(n_+<0)$ are well suited for adiabatic (diabatic) evolution. Concretely, we plot as solid (dashed) lines the diabatic (adiabatic) evolution in terms of the energy diagram (Fig.~\ref{fig: LZ}a), the di-ad metric (Fig.~\ref{fig: LZ}c) yields the corresponding pulse shapes (Fig.~\ref{fig: LZ}b) and infidelities (Fig.~\ref{fig: LZ}d).
The highest diabatic state transfer ($\ket{\psi_0}\mapsto \ket{\psi_1}$) fidelities are acquired for $x \,t_\text{f}\rightarrow 0$ (See Supplemental Material~\ref{app: sudden approximation}). In essence, the resulting state transfer fidelity can be maximized by maximizing the energy fluctuations around the smallest energy gap, which in our protocol corresponds to choosing $\hat{\alpha},\hat{\beta}<0$. In contrast, and also in light of the quantum metric tensor and Eq.~\eqref{eqn: energy fluctuations qmt}, pulses with $\alpha,\beta>0$ allow for possibly reduced energy uncertainty around the energy gap. Due to the flexibility of adjusting the adiabatic and diabatic parameters $(\alpha,\beta;\, \hat{\alpha},\hat{\beta})$ simultaneously, we can not only allow for purely diabatic or adiabatic transfer protocols (by maximizing or minimizing the energy fluctuations), but we can also generate pulses that interpolate between these two, providing the ability to maintain adiabatic evolution for some level crossings whilst being diabatic for others.

\begin{figure}[t!]
    \centering
    \includegraphics[width=\linewidth]{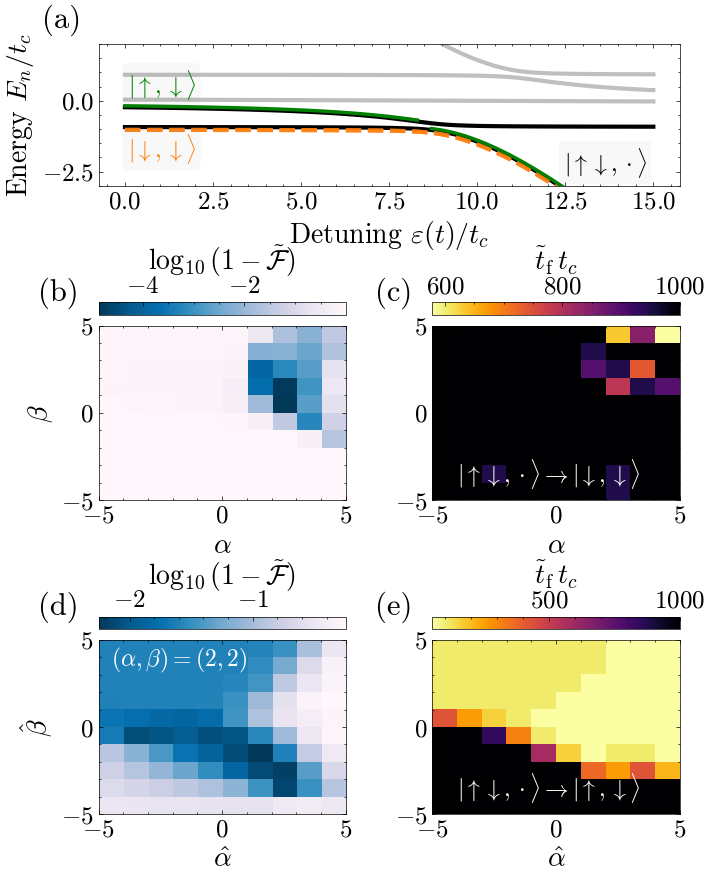}
    \caption{Initialization of a spin qubit in a DQD. The energy spectrum is shown \textbf{(a)}. The optimal infidelity and corresponding pulse times for the adiabatic protocol ($\ket{\uparrow\downarrow,\cdot}\mapsto \ket{\downarrow,\downarrow}$) are illustrated in \textbf{(b),(c)}. Similarly, for the di-ad protocol  ($\ket{\uparrow\downarrow,\cdot}\mapsto \ket{\uparrow,\downarrow}$), we showcase the infidelities and pulse times in \textbf{(d),(e)}. The DQD parameters read $\tilde{U}/t_c=10, \,E_Z/t_c=0.9,\, \Delta\!E_Z/t_c=0.1, \,\Delta\!E_X/t_c=0.01$, where we sweep the detuning $\varepsilon(0)/t_c=15$ to $\varepsilon(t_\text{f})/t_c=0$. For the di-ad protocol, we fix the adiabatic parameters to $(\alpha,\beta){=}(2,2)$. For the protocol, we sweep possible pulse times $t_\text{f}t_c\in [0,1000]$.}
    \label{fig: dqd init}
\end{figure}

\textit{Versatile state initialization --- }To demonstrates our protocol, we enhance the initialization of a semiconductor spin qubit~\cite{burkardSemiconductorSpinQubits2023} within a double quantum dot (DQD) using spin-to-charge conversion~\cite{seedhousePauliBlockadeSilicon2021, niegemannParitySingletTripletHighFidelity2022, zhangUniversalControlFour2024, jirovecSinglettripletHoleSpin2021, jirovecDynamicsHoleSingletTriplet2022, fernandez-fernandezQuantumControlHole2022, fernandez-fernandezLongrangeTransportAsymmetric2025, tidjaniThreeDimensionalArrayQuantum2025, jirovecMitigationExchangeCrosstalk2025, farinaSiteresolvedMagnonTriplon2025}. A common bottleneck for high-fidelity initialization is the emergence of small gaps caused by small Zeeman vector differences~\cite{farinaSiteresolvedMagnonTriplon2025}. The di-ad tensor enables a combination of adiabatic and diabatic dynamics through the transition matrix $\xi_{mn}$. This flexibility enables us to develop state transfer protocols tailored to versatile spin-to-charge conversion~\cite{rimbach-russGaplessSingleSpinQubit2025, kellyIdentifyingMitigatingErrors2025}. The spin-to-charge conversion inside a DQD system is well-described by an extended Fermi-Hubbard-Zeeman Hamiltonian~\cite{sm}. Projecting the model to the relevant charge states $(n_L,n_R){=}(2,0)$ and $(1,1)$ with the associated spin basis states $\{\ket{\uparrow\downarrow,\cdot},\ket{\uparrow,\uparrow},\ket{\uparrow,\downarrow},\ket{\downarrow,\uparrow},\ket{\downarrow,\downarrow}\}$, results in the following matrix representation~\cite{ungererStrongCouplingMicrowave2024, geyerAnisotropicExchangeInteraction2024, kellyIdentifyingMitigatingErrors2025}
    \begin{align}
\label{eqn: 5x5 Hamiltonian}
    \hat{H}_{\text{DQD}} = \begin{pmatrix}
 \tilde{U}{-}\varepsilon(t) & 0 & -t_c & t_c & 0 \\
 0 & E_Z & \Delta\!E_X & -\Delta\!E_X & 0 \\
 -t_c & \Delta\!E_X & \Delta\!E_Z & 0 & \Delta\!E_X \\
 t_c & -\Delta\!E_X & 0 & -\Delta\!E_Z & -\Delta\!E_X \\
 0 & 0 & \Delta\!E_X & -\Delta\!E_X & -E_Z
\end{pmatrix}.
\end{align}
Here $t_c$ is the tunnel coupling, $E_a=E_{a,1}+E_{a,2}$ and $\Delta\!E_a=E_{a,1}-E_{a,2}$ are the total Zeeman energy and the Zeeman splitting difference (with $a=x,y,z$) arising from a spatially varying g-factor \cite{vanriggelen-doelmanCoherentSpinQubit2024, vanriggelenPhaseFlipCode2022, boscoFullyTunableLongitudinal2022, hsiaoExcitonTransportGermanium2024, geyerAnisotropicExchangeInteraction2024, saez-mollejoExchangeAnisotropiesMicrowavedriven2025, nguyenSinglestepHighfidelityThreequbit2025, jakobFastReadoutQuantum2025, valvoElectricallyTuneableVariability2025} or a magnetic field gradient \cite{ takedaQuantumErrorCorrection2022, xueQuantumLogicSpin2022, desmetHighfidelitySinglespinShuttling2025}.  In addition, we define an effective Coulomb repulsion $\Tilde{U}$ and the detuning $\varepsilon\in \mathbb{R}_{\geq 0}$ between the DQD chemical potentials. For spin-qubit initialization we employ the spin-to-charge conversion method, which requires the pulse shaping of the detuning $\varepsilon(t)$ from a far detuned position $\varepsilon(0)\gg |t_c|, |E_Z|$, where the ground state is given by two charges in one quantum dot (i.e., $\ket{\uparrow\downarrow, \cdot}$), to the symmetric point $\varepsilon(t_\text{f})=0$, where we have transferred one charge to the other dot $(2,0){\to}(1,1)$. Thereby, we initialize a computational state. Due to a finite spin-flipping term $\Delta\!E_X$, which causes a small anticrossing between the ground and first excited state, a spin-flip event may occur during spin-to-charge conversion (See Fig.~\ref{fig: dqd init}a). We can make use of this hybridization to initialize two sets of computational states via our di-ad method. Starting from two charges in one dot, we have two initialization schemes, resulting in different parity sectors:
\begin{itemize}
    \item Adiabatic evolution: $\ket{\uparrow\downarrow,\cdot}\mapsto \ket{\downarrow,\downarrow}$ 
    \item Diabatic-adiabatic evolution: $\ket{\uparrow\downarrow,\cdot}\mapsto \ket{\uparrow,\downarrow}$
\end{itemize}
\begin{figure*}
    \centering
    \includegraphics[width=0.96\textwidth]{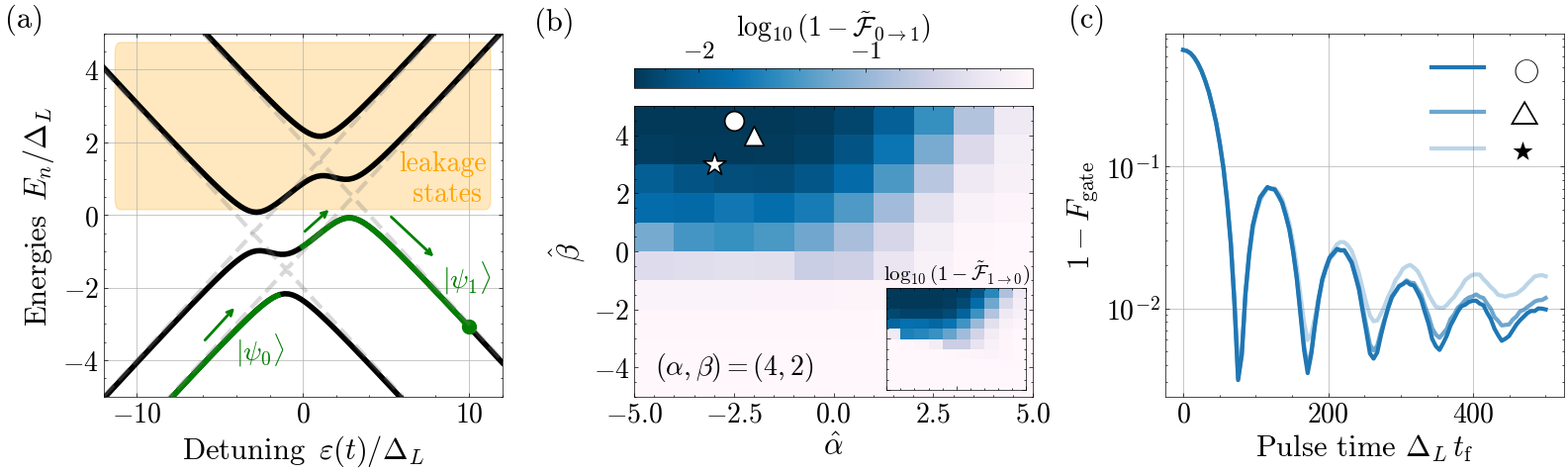}
    \caption{On-the-go operations using shuttling-induced excitations. The energy spectrum of the Hamiltonian~\eqref{eqn: 4x4 Ham}, where we marked in orange the undesired leakage states, is plotted in \textbf{(a)}. In addition, we plot the minimum infidelities $1-\tilde{\mathcal{F}}$ for the state transfers $\ket{\psi_0}{\to}\ket{\psi_1}$ \textbf{(b)} and $\ket{\psi_1}{\to}\ket{\psi_0}$ (inset), for possible pulse times   $\Delta_Lt_\text{f}\in [0,500]$. Using the Horodecki identity, in \textbf{(c)}, we analyze the effective X-gate average gate fidelity $F_\text{gate}=(1+\sum_{j=0,1} |\braket{\psi_j(0)}{\psi_{1-j}(t_\text{f})}|^2)/3$ for the computational subspace spanned by the lower two eigenstates as a function of possible pulse times $\Delta_L\,t_\text{f}$ for the indicated circle, square and star parameters in \textbf{(b)}. The parameters used are $t_c/\Delta_L=.1,\Delta_R/\Delta_L=5, \phi_L=0.1, \phi_R=\pi/2$. The detuning is swept from $\varepsilon/\Delta_L\in[-10,10]$.}
    \label{fig: shuttling}
\end{figure*}
For the diabatic-adiabatic case, we have the following di-ad transition matrix (in the eigenbasis of the Hamiltonian) that is fully determined by the initial and final states
\begin{align}
    \xi_{mn}\big(\ket{\uparrow\downarrow,\cdot}\mapsto \ket{\uparrow,\downarrow}\big)=\begin{pmatrix}
\cdot & 1 & 0 & 0 & 0 \\
1 & \cdot & 0 & 0 & 0 \\
0 & 0 & \cdot & 0 & 0 \\
0 & 0 & 0 & \cdot & 0 \\
0 & 0 & 0 & 0 & \cdot
\end{pmatrix}.
\end{align}
The dots represent the excluded transitions. In Fig.~\ref{fig: dqd init} we simulate both initialization procedures, as shown in Fig.~\ref{fig: dqd init}a), and plot the minimum infidelity $1{-}\tilde{\mathcal{F}}\equiv 1{-}\mathcal{F}(\tilde{t}_\text{f})$ (b, d) and the corresponding (rescaled) pulse time $\tilde{t}_\text{f}\,t_c$ (c, e). For the purely adiabatic case (Fig.~\ref{fig: dqd init}b, c), we only require sweeping the adiabatic parameters $(\alpha,\beta)$ as we aim to suppress excitations to any of the 4 possible excited states. For the diabatic-adiabatic initialization (Fig.~\ref{fig: dqd init}d, e), we fix the adiabatic parameters $(\alpha,\beta){=}(2,2)$ and sweep the diabatic ones $(\hat{\alpha}, \hat{\beta})$. Remarkably, both methods achieve fidelities exceeding $99\%$ with short simulation times~\cite{sm}. We can further enhance the efficiency of the di-ad protocol by using machine learning sampling methods to obtain the desired di-ad parameter configuration with even fewer infidelity simulations~\cite{sm}. By allowing for the interpolation of diabatic and adiabatic dynamics, we unlock the ability to initialize different parity sectors, which, for instance, allows for flexible initial qubit configurations in a quantum circuit while reducing the circuit overhead for similar initial configurations or for analog simulation of many-body systems with limited individual qubit control~\cite{farinaSiteresolvedMagnonTriplon2025, jirovecManybodyInterferometrySemiconductor2025}.

\textit{Operations during shuttling --- }Another state-transfer protocol important for quantum information processing is shuttling~\cite{ginzelSpinShuttlingSilicon2020, pazhedathLargeSpinshuttlingOscillations2025, nemethOmnidirectionalShuttlingAvoid2024, desmetHighfidelitySinglespinShuttling2025, losertStrategiesEnhancingSpinShuttling2024, fernandez-fernandezSpinorbitenabledRealizationArbitrary2025, matsumotoTwoqubitLogicTeleportation2025, ademiDistributingEntanglementDistant2025}, where the qubit is physically moved around the quantum chip to enable sparse quantum architectures to minimize crosstalk and provide opportunities to include classical electronics on-chip~\cite{vandersypenInterfacingSpinQubits2017, tosatoCrossbarChipBenchmarking2026, liTrilinearQuantumDot2025}. Furthermore, optimized bucket brigade shuttling is necessary for the operation of sparse, single-barrier control spin-qubit architectures~\cite{ivlevOperatingSemiconductorQubits2025}. For silicon quantum dots, we can model the Hilbert space by a tensor product structure $\mathcal{H}=\mathcal{H}_\text{orbital}\otimes \mathcal{H}_\text{valley}$, whose Hamiltonian, after a unitary transformation, is given by~\cite{nemethOmnidirectionalShuttlingAvoid2024, thayilTheoryValleySplitting2025}
\begin{align}
\label{eqn: 4x4 Ham}
    \hat{H}_\text{BB}
    &=\begin{pmatrix}
        \varepsilon(t) + |\Delta_L| & 0 & t_{ee} & t_{eg} \\
        0 & \varepsilon(t)-|\Delta_L| & t_{ge} & t_{gg} \\
        t_{ee}^* & t_{ge}^* &  |\Delta_R| & 0 \\
        t_{eg}^* & t_{gg}^* & 0 &  - |\Delta_R| 
    \end{pmatrix},
\end{align}
where, as before the detuning is given by $\varepsilon(t)$ and will constitute our control parameter, $\Delta_{L,R}\in \mathbb{C}$ are the valley terms resulting in the valley splitting $E_\text{VS}=2\abs{\Delta}$, and the bare tunnel coupling $t_c$ is modified by the valley phases $\phi_{L/R}$
\begin{align}
    t_{ee}&=t_{gg}^*=\frac{t_c}{2}\left(1+e^{i(\phi_L-\phi_R)}\right),\\
    t_{eg}&=-t_{ge}^*=\frac{t_c}{2}\left(e^{i\phi_L}-e^{i\phi_R}\right).
\end{align}
We note that a similar Hamiltonian emerges in the shuttling of holes in germanium quantum dots by changing to a different Hilbert space decomposition, namely $\mathcal{H}=\mathcal{H}_\text{orbital}\otimes \mathcal{H}_\text{spin}$~\cite{wangOperatingSemiconductorQuantum2024}. The two lowest energy eigenstates of the Hamiltonian~\eqref{eqn: 4x4 Ham} will constitute our computational subspace as seen in Fig.~\ref{fig: shuttling}a. Using our di-ad tensor, we can perform two relevant operations, namely, while shuttling or hopping, we can excite the ground state to the first excited state, while suppressing leakage (See Fig.~\ref{fig: shuttling}b), and we can also force the first excited state to relax to the ground state (See inset of Fig.~\ref{fig: shuttling}b). Both shuttling operations achieve state-transfer fidelities of 99\%, alongside an average effective X-gate fidelity with similar performance (See Fig.~\ref{fig: shuttling}c), demonstrating the flexibility, scalability, and relevance for current quantum information processing.

\textit{Conclusion --- }We developed a general method for optimal state-transfer that is capable of minimizing undesired transitions, while allowing for targeted excitations to allow for full navigation of any multi-level energy landscape. In addition, the single-parameter di-ad protocols are efficiently computable, as the optimization task reduces to a first-order differential equation. Furthermore, the inclusion of the parameters $(\alpha,\beta;\, \hat{\alpha},\hat{\beta})$ enables consideration of hardware limitations by design, making it highly adaptable to current experimental constraints.

\textit{Acknowledgments --- }We thank David Fernandez-Fernandez, Kristof Moors, George Simion, Yuta Matsumoto, Sander de Snoo, and all members of the Rimbach-Russ and Bosco group for providing valuable insights.

\textit{Funding --- }M.R.-R., S.B., E.V., and C.V.M. acknowledge that the EU partly supported this research through the H2024 QLSI2 project and was partly sponsored by the Army Research Office under Award Number: W911NF-23-1-0110. M.R.-R. and E.V. acknowledge support from the Dutch Research Council (NWO) under Award Number Vidi TTW 22204. The views and conclusions contained in this document are those of the authors and should not be interpreted as representing the official policies, either expressed or implied, of the Army Research Office or the U.S. Government. The U.S. Government is authorized to reproduce and distribute reprints for Government purposes, notwithstanding any copyright notation herein.

\textit{Data availability --- }The data that support the findings of this article are openly available at~\footnote{Public repository found at \url{https://doi.org/10.5281/zenodo.18612868}.}

\textit{Author contributions --- }C.V.M. and E.V. performed the theoretical computations and numerical simulations with inputs from M.R.-R. and S.B.. C.V.M. wrote the manuscript with inputs from all the authors.

\newpage
\begin{widetext}
    \begin{center}
    \textbf{Supplemental Material: Multi-level spectral navigation with geometric diabatic-adiabatic control}\newline 
    \text{\small Christian Ventura-Meinersen, Edmondo Valvo, Stefano Bosco, Maximilian Rimbach-Russ}

    \vspace{0.5cm}
    In this supplemental material, we discuss the motivation of the diabatic evolution ansatz, describe the pulse generation, and derive the double quantum dot model. Finally we also investigate the pulse generation simulation runtime and also investigate the use of automated paramter optimization strategies.
\end{center}
\end{widetext}

\section{Ansatz for diabatic control}
\label{app: sudden approximation}
We start from time-dependent perturbation theory, where the time-dependent coefficients of the wave function, in the eigenbasis expansion, read
\begin{align}
\begin{split}
    \dot{c}_m=&-c_m\langle\psi_m|\dot{\psi}_m\rangle\\
    &-\sum_{n\neq m}c_n\frac{\mel{\psi_m}{\dot{H}}{\psi_n}}{E_n-E_m}e^{-i\int_0^t (E_n-E_m) dt'}.
    \label{eq:time_evolution}
\end{split}
\end{align}
In the impulse approximation~\cite{tomkaAccuracyAdiabaticimpulseApproximation2018, suzukiGeneralizedAdiabaticImpulse2022}, we can simplify $\langle\psi_m|\dot{\psi}_m\rangle \approx 0$, such that the above expression reduces to
\begin{align}
    \dot{c}_m&\approx -\sum_{n\neq m} c_n\frac{\mel{\psi_n}{\dot{H}}{\psi_m}}{E_m-E_n} \equiv  -\sum_{n\neq m} c_n f_{nm}.
\end{align}
With this, we can compute the fidelity of leaving the initial state by approximating the integral around $t_\text{f}$
\begin{align}
    c_m(t_\text{f})\approx c_m(0) - t_\text{f}\sum_{n\neq m}c_n(t_\text{f})f_{nm}(t_\text{f}).
\end{align}
The transfer fidelity is given by
\begin{align}
    \mathcal{F}^{(\neq m)}_\text{di}&=1-\abs{c_m(t_\text{f})}^2\\ \nonumber
    &= 1 - \Big(|c_m(0)|^2-t_\text{f}c_m^*(0)\sum_{n\neq m} c_n(t_\text{f}) f_{nm}(t_\text{f})\\ \nonumber
    &-t_\text{f}c_m(0)\sum_{n\neq m} c_n^*(t_\text{f}) f_{nm}^*(t_\text{f})+\mathcal{O}(t_\text{f}^2)\Big) \nonumber
\end{align}
where, if we start in the $m$-th eigenstate we have $c_m(0){=}1$, results in (in linear order of $t_\text{f}$)
\begin{align}
    \mathcal{F}^{(\neq m)}_\text{di}\approx 2t_\text{f}\Re\sum_{n\neq m} c_n(t_\text{f}) f_{nm}(t_\text{f}).
\end{align}
To maximize the fidelity of leaving the initial state $\ket{\psi_m}$, we need to maximize the sum on the right-hand side. In our expansion, we assumed a small and non-negative right-hand side. This allows us to maximize the transfer fidelity
\begin{align}
    \max\mathcal{F}^{(\neq m)}_\text{di}&\simeq \max\left(\sum_{n\neq m} c_n(t_\text{f}) f_{nm}(t_\text{f})\right).
\end{align}
For a two-level system, where we aim to find $c_{m=0}(t_\text{f}){=}0$ and $c_{n=1}(t_\text{f}){=}1$, we find that
\begin{align}
    \max\mathcal{F}^{(\neq m)}_\text{di}\simeq \max f_{01}(t_\text{f}) = \min \frac{E_1-E_0}{\mel{\psi_0}{\dot{H}}{\psi_1}}.
\end{align}
These results motivate our finding that, for $\hat{\alpha},\hat{\beta}<0$, high-fidelity diabatic protocols are obtained. 

\section{Pulse shape generation}
\label{app: pulse shape generation}
\begin{figure}
    \centering
    \includegraphics[width=0.8\linewidth]{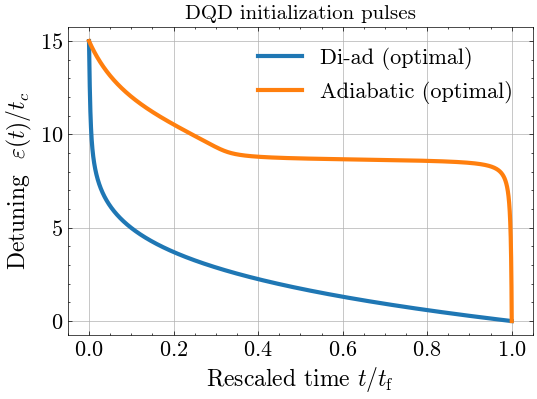}
    \caption{Pulse shapes for highest-fidelity DQD initialization using the di-ad protocol (blue) and fully adiabatic (orange) in Figure~\ref{fig: dqd init}.}
    \label{fig: dqd init pulses}
\end{figure}

State transfer protocols are given by a path connecting the set of initial parameter values $x^\mu_\text{i}\equiv x^\mu(0)$ to some final set $x^\mu_\text{f}\equiv x^\mu(t_\text{f})$. The task of fast and high-fidelity population transfer relates to the optimization problem of finding an optimal path $x^\mu_\text{geo}(t)$ between these two points described by the following functional
\begin{align}
    \mathcal{L}[x,\dot{x}]=\int_{0}^{t_\text{f}} dt \,\sqrt{g_{\mu \nu}(x)\dv{x^\mu}{t}\dv{x^\nu}{t}}.
\end{align}
This functional describes the length of a path $x^\mu(t)$ parametrized by time $t$. Due to the parametrization invariance, we can choose any affine parameter $t\to \xi t$ with $\xi \in \mathbb{R}$. This integral expression can be minimized for functions $x^\mu_\text{geo}(t)$ that solve the Euler-Lagrange equations, which in this context are known as the geodesic equations. We can reformulate the geodesic equations as~\cite{ventura-meinersenQuantumGeometricProtocols2025}
\begin{align}
    \label{eqn: adiabatic quantum geo protocol}
    g_{\mu \nu}(x)\dv{x^\mu}{t}\dv{x^\nu}{t}=\delta^2\ll 1.
\end{align}
For now, $\delta$ is a constant parameter ensuring the boundary conditions for the pulse. As $\delta^2\propto\sigma_E^2$, we call $\delta$ the adiabaticity since it is proportional to the energy uncertainty. If we restrict ourselves to a single parameter $x^\mu=\varepsilon(t)$, we can solve for the adiabaticity parameter as follows
\begin{align}
    \label{eqn: adiabaticity defintion}
    \delta = \frac{1}{t_\text{f}}\int_{\varepsilon(0)}^{\varepsilon(t_\text{f})}d\varepsilon \, \sqrt{g_{\varepsilon \varepsilon}}=\frac{\mathcal{L}[\varepsilon]}{t_\text{f}}\ll 1.
\end{align}
Therefore, adiabatic protocols can be understood as paths that minimize locally the length of the path that they trace out, i.e. short geodesics with respect to the time $t_\text{f}$. In~\cite{meinersenUnifyingAdiabaticStatetransfer2025}, it was shown that a similar procedure can be done by simply substituting the quantum metric tensor $g_{\mu\nu}$ with an extended version. Here, we will continue this trend by substituting $g_{\mu\nu}\mapsto \mathcal{G}_\text{di-ad}$, allowing for an interpolation between adiabatic and diabatic dynamics. Concretely, for a single parameter $x^\mu=\varepsilon(t)$, we need to solve the following ordinary differential equation 
\begin{align}
    \Big(\mathcal{G}_\text{di-ad}\Big)_{\varepsilon\varepsilon}^{(\alpha,\beta;\hat{\alpha},\hat{\beta})}\left(\dv{\varepsilon(t)}{t}\right)^2=\delta^2,
\end{align}
with the constant $\delta$ given by
\begin{align}
    \delta = \frac{1}{t_\text{f}}\int_{\varepsilon(0)}^{\varepsilon(t_\text{f})}d\varepsilon \, \sqrt{\Big(\mathcal{G}_\text{di-ad}\Big)_{\varepsilon\varepsilon}^{(\alpha,\beta;\hat{\alpha},\hat{\beta})}}.
\end{align}
In Figure~\ref{fig: dqd init pulses}, we illustrate the resulting pulse shapes for the highest fidelity  DQD initialization.
We may wish to study a few-parameter control to accomplish a specific state transfer protocol. In this case, the full geodesic equations are needed. For a given metric $g_{\mu\nu}$ and parameters $x^\mu$, the geodesic equations read
\begin{align}
    \dv[2]{x^\mu}{\tau}+\Gamma^\mu_{\alpha,\beta} \dv{x^\alpha}{\tau}\dv{x^\beta}{\tau}=0.
\end{align}
Here $\Gamma^\mu_{\alpha,\beta}$ represent the Christoffel symbol and are defined as
\begin{align}
    \Gamma^\mu_{\alpha,\beta}=\frac{1}{2}\, g^{\mu\nu}
\left(
\frac{\partial g_{\nu\alpha}}{\partial x^\beta}
+ \frac{\partial g_{\nu\beta}}{\partial x^\alpha}
- \frac{\partial g_{\alpha\beta}}{\partial x^\nu}
\right).
\end{align}

\section{Double quantum dot model}
\label{app: DQD model}
In the main text, we derived an effective model for the DQD, which can be derived using the Fermi-Hubbard-Zeeman Hamiltonian 
\begin{multline}
\hat{H}_\text{FM-Z} = -t_c\sum_{ij,\sigma}\Big(\hat{c}_{i,\sigma}^\dagger \hat{c}_{j,\sigma} +\text{h.c.}\Big) +\sum_{\langle ij\rangle}U_{ij}\hat{n}_i\hat{n}_j\\
+\sum_j \left(\frac{U}{2}\hat{n}_j(\hat{n}_j-1)+V_j\hat{n}_j\right)+\frac{1}{2}\mu_B \sum_j \vec{\mathcal{B}}^j \cdot \vec{\sigma}^j
\end{multline}
where $\hat{c}^\dagger_{j,\sigma}(\hat{c}_{j,\sigma})$ creates (annihilates) a fermion on sites $j$ with spin $\sigma$. The fermionic number operator is $\hat{n}_j=\sum_\sigma \hat{c}^\dagger_{\sigma,j}\hat{c}_{\sigma,j}$, $U$ and $U_{ij}$ are the intra- and inter-dot Coulomb repulsion, $t_c$ is the tunnel coupling, and $V_j$ are the chemical potentials in each dot. The spin degeneracy is lifted through the Zeeman term,
where $\mu_B$ is the Bohr magneton, $\vec{\sigma}=(\sigma_x,\sigma_y,\sigma_z)^T$ is the Pauli vector consisting of the Pauli matrices, $\mathcal{B}^j_a=\sum_b g^j_{ab} B_b$ the effective magnetic field and $g^j_{ab}$ the g-tensor at site $j$ with $a, b \in \{x,y,z\}$. The effective Coulomb repulsion is given by $\Tilde{U}=U{-}U_{12}$ and the detuning is $\varepsilon=V_2{-}V_1$.

\section{Runtime analysis}
\label{app: runtime analysis}
\begin{figure}
    \centering
    \includegraphics[width=\linewidth]{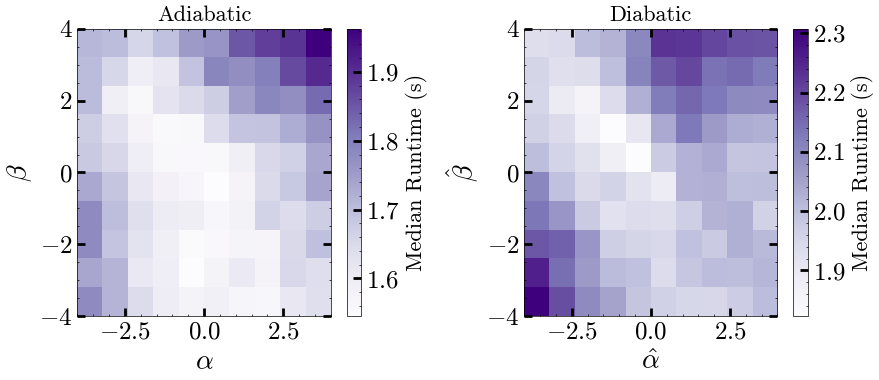}
    \caption{Runtime analysis of the pulse generation for the versatile DQD initialization as a function of the same parameters as in Figure~\ref{fig: LZ}. We take the median of 100 runs to average out CPU fluctuations.}
    \label{fig: runtime}
\end{figure}

In this section, we investigate the resulting runtime it takes to simulate a given pulse for a combination of $(\alpha,\beta;\,\hat{\alpha},\hat{\beta})$. We focus on the DQD initialization as a realistic application of our model. The results are shown in Figure~\ref{fig: runtime}. Notably, the runtime does depend on the values of the parameters $(\alpha,\beta;\,\hat{\alpha},\hat{\beta})$. Importantly, we only require the simulation of a single pulse for the combinations of parameters. We can use the same pulse for different total pulse times $t_\text{f}$. This is granted by the reparametrization invariance of our resulting di-ad protocol.  
\section{Automated parameter optimization}
\label{app: sampling ML}

\begin{figure}
    \centering
    \includegraphics[width=0.8\linewidth]{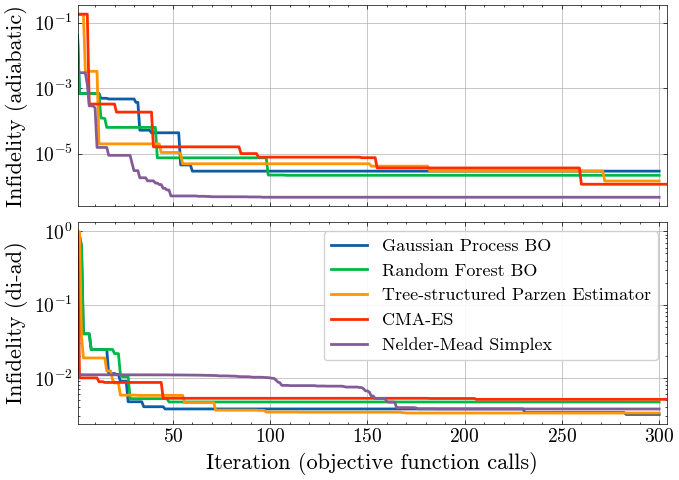}
    \caption{DQD di-ad parameter optimization for the adiabatic (top) and di-ad (bottom) initialization strategies. All parameters used are the same as in the main text, and the pulse time is fixed to $t_\text{f}t_c=500$. The used bounds are $\alpha,\beta \in [0,5]$ and $\hat{\alpha},\hat{\beta}\in [-3,3]$.}
    \label{fig: sampling ML}
\end{figure}

The simulation speed of the di-ad pulses is aided by the fact that we can perturbatively compute corrections to the pulse by neglecting higher-order contributions. In addition, the pulses are time reparametrization invariant, and hence for a single combination of parameters, one can use the pulse for all possible pulse times $t_\text{f}$. Therefore, the only bottleneck in confirming which di-ad parameters are optimal is the computation of the infidelity. To avoid high-cost mappings, we use the following optimizers from the \texttt{skopt}~\cite{headScikitoptimize2021}, \texttt{optuna}~\cite{akibaOptunaNextgenerationHyperparameter2019}, \texttt{cma}~\cite{hansenCMAEvolutionStrategy2023}, \texttt{scipy}~\cite{virtanenSciPy10Fundamental2020} Python libraries to find the optimal solution:
\begin{itemize}
    \item Gaussian Process Bayesian Optimization
    \item Random Forest Bayesian Optimization
    \item Tree-structured Parzen Estimators
    \item Covariant Matrix Adaptation Evolution Strategy 
    \item Nelder-Mead Simplex Method.
\end{itemize}
All used methods attempt to approximately optimize a given cost function with parameters $\theta$
\begin{align}
    \theta^*=\text{argmin}_{\theta \in \mathbb{R}}[1-\mathcal{F}(\theta)].
\end{align}
where $\theta=(\alpha,\beta,\hat{\alpha},\hat{\beta})$ represents our parameters to optimize. In Figure~\ref{fig: sampling ML}, we investigate these optimization strategies for the DQD system in Eq.~\eqref{eqn: 5x5 Hamiltonian} for a fixed pulse time. 
For both initialization strategies, we find the corresponding optimal solution in a few objective function evaluations.

\newpage

\bibliography{references}

\end{document}